\documentclass[journal]{IEEEtran}
\usepackage{verbatim}
\usepackage{graphicx}
\usepackage{xcolor,cite}
\usepackage[normalem]{ulem}
\usepackage[cmex10]{amsmath}
\usepackage{amssymb}
\usepackage{array}
\usepackage{mdwmath}
\usepackage{mdwtab}
\usepackage{eqparbox}
\usepackage{url}
\usepackage[export]{adjustbox}
\usepackage{kantlipsum}

\begin{document}
\title{Broad-Angle Multichannel Metagrating Diffusers}

\author{Yarden~Yashno,~\IEEEmembership{Student Member,~IEEE,}
        and~Ariel~Epstein,~\IEEEmembership{Senior Member,~IEEE} 
\thanks{The authors are with the Andrew and Erna Viterbi Faculty of Electrical and Computer Engineering, Technion - Israel Institute of Technology, Haifa 3200003, Israel (e-mail: epsteina@ee.technion.ac.il).}%
\thanks{Manuscript received August XX, 2022; revised ***, **, ****.}}
\markboth{IEEE Transactions on Antennas and Propagation,~Vol.~**, No.~*, Month~****}%
{Yashno and Epstein: Multichannel MG Diffusers}
\maketitle

\begin{abstract}
We present a semianalytical scheme for the design of broad-angle multichannel metagratings (MG), sparse periodic arrangements of loaded conducting strips (meta-atoms), embedded in a multilayer printed circuit board configuration. By judicious choice of periodicity and angles of incidence, scattering off such a MG can be described via a multi-port network, where the input and output ports correspond to different illumination and reflection directions associated with the same set of propagating Floquet-Bloch modes. Since each of these possible scattering scenarios can be modelled analytically, constraints can be conveniently applied on the modal reflection coefficients (scattering matrix entries) to yield a diffusive response, which, when resolved, produce the required MG geometry. We show that by demanding a symmetric MG configuration, the number of independent S parameters can be dramatically reduced, enabling satisfaction of \emph{multiple} such constraints using a \emph{single} sparse MG. Without any full-wave optimization, this procedure results in a fabrication-ready layout of a multichannel MG, enabling retroreflection suppression and diffusive scattering from \emph{numerous} angles of incidence \emph{simultaneously}. This concept, verified experimentally via a five-channel prototype, offers an innovative solution to both monostatic and bistatic radar cross section reduction, avoiding design and implementation challenges associated with dense metasurfaces used for this purpose.
\end{abstract}


\IEEEpeerreviewmaketitle

\section{Introduction} \label{sec:introduction}
\IEEEPARstart{T}{he} radar cross section (RCS) of an object is a measure for the ability to detect it by the power scattered {off it} in a given direction when illuminated by a radar \cite{E.F.Knott2004RadarSection}. {Since controlling such scattering is crucial for} many {defense} applications, {means for} RCS {reduction} ha{ve} been studied extensively over the years. The RCS depends on the {respective positions of the radar antennas and} the scatterer; it is {customary} to distinguish between bistatic radar scenarios, where {the} transmitter and receiver are separate{d} by a distance comparable to the{ir} distance {to the target}, and monostatic radar scenarios, where the transmitter and receiver are co-located. {While in the former case, the angle of incidence upon the target generally differs from the reflection angle that would be intercepted by the receiver,} in the latter, the detection is based on the retroreflection {from {the target}}. Therefore, {and considering that the angle of illumination is unknown in general, {the ideal} RCS reduction cover would need to diminish scattered fields in as many directions as possible, and perform equally well for multiple excitation scenarios.}

In recent years, metasurfaces (MS), thin sheets of closely-packed subwavelength {polarizable particles} (meta-atoms), have been {extensively investigated as means to tackle such challenges}  \cite{Glybovski2016Metasurfaces:Visible}. Being low-profile and conformal, {and with their proven ability to control wave scattering for various applications,} they {have} the potential to {serve as {effective} target} coatings{, acting} to reduce their radar signature. One approach to achieve this is {to design absorbing MSs} {\cite{Radi2015ThinRealizations, Lim2019UltrawidebandMetasurfaces}}, {which could be used to suppress} both monostatic and bistatic RCS {at the same time}. However, {such devices} require engineering lossy materials within thin sheets, which is {often} nontrivial from a practical standpoint \cite{Murugesan2021ATechniques}.

Alternatively, one can avoid such complexities while still obtaining substantial RCS reduction with ultrathin covers by using diffusive MSs. One {such} common {design relies on the} the checkerboard {configuration}, as {proposed} in {\cite{Paquay2007ThinReduction, IriarteGalarregui2013BroadbandTechnology,Ghayekhloo2018AnModel}}{.} 
These structures are composed of {alternating} perfect electric conductor (PEC) and artificial magnetic conductor (AMC) {reflective unit cells; when excited by a} 
normal{ly} incident plane wave, the {$\pi$ phase shift between the PEC and AMC reflected waves} caus{e} destructive interference in that direction. 
While these {surfaces} are relatively simple to design and manufacture, they {aim} to reduce the monostatic RCS for normal{-direction} excitation scenarios (where the specular reflection coincides with the retroreflection), yet expose the target for detection from {other} angles. 

\textcolor{black}{An {evolution} of this idea, presented in \cite{Chen2014ReductionMetasurface, Cui2014CodingMetamaterials,Gao2015BroadbandMetasurfaces,Chen2016GeometricScattering,Moccia2017CodingDesign,Feng2018Two-dimensionalReduction, Azizi2020Ultra-widebandSurface}, uses diffusive metasurfaces composed of elements exhibiting varying reflection phases, {thereby} {purposely} deflecting the specular reflection towards {(one or more)} different directions.} However, since the reflection phases correspond to normal{ly} inciden{t wave excitations}, wide-angle response is generally not guaranteed. Extending the angular response is typically obtained via brute-force optimization methods, not as an inherent part of the design, which is usually time consuming and lacks physical insight \cite{Murugesan2021ATechniques}. 

Further to the challenges described in the previous paragraph{s}, utilizing MS{s} for RCS reduction faces an additional, more fundamental difficulty, resulting from the need to obey the homogenization approximation. To form the equivalent surface constituents\footnote{{In the {aforementioned} \cite{Paquay2007ThinReduction, IriarteGalarregui2013BroadbandTechnology,Chen2014ReductionMetasurface, Cui2014CodingMetamaterials, Gao2015BroadbandMetasurfaces,Chen2016GeometricScattering, Moccia2017CodingDesign,Ghayekhloo2018AnModel,Feng2018Two-dimensionalReduction, Azizi2020Ultra-widebandSurface}, these constituents are effectively the local reflection coefficient magnitudes and phases.}} yielding the generalized sheet transition conditions (GSTC) {governing most of the synthesis procedures} \cite{Kuester2003AveragedMetafilm}{, dense closely-packed meta-atom arrangements should be used}. {Such a requirement, however, tends to} complicat{e} the design and fabrication process{es} {due to the need to devise and realize a large number of particles with deep subwavelength features and small separation distances}. In addition, following the typical GSTC-oriented synthesis approach results in abstract MS constituent distribution{, which generally implements the desired surface response for a given (single) excitation\footnote{The designated excitation is typically a normally incident plane wave \cite{Paquay2007ThinReduction, IriarteGalarregui2013BroadbandTechnology,Chen2014ReductionMetasurface, Cui2014CodingMetamaterials, Gao2015BroadbandMetasurfaces,Chen2016GeometricScattering, Moccia2017CodingDesign,Ghayekhloo2018AnModel,Feng2018Two-dimensionalReduction, Azizi2020Ultra-widebandSurface}.}\cite{Epstein2016HuygensApplications}. Thus, even if one succeeds translating these constituents into a practical MS prototype, usually via full-wave simulations utilized to design the physical geometries associated with the various meta-atoms, attributing multifunctional properties to the MS (e.g., accommodating multiple angles of incidence) forms yet another nontrivial challenge.}

\begin{figure*}[htbp]
\centering
\includegraphics[width=\textwidth,height=\textheight,keepaspectratio]{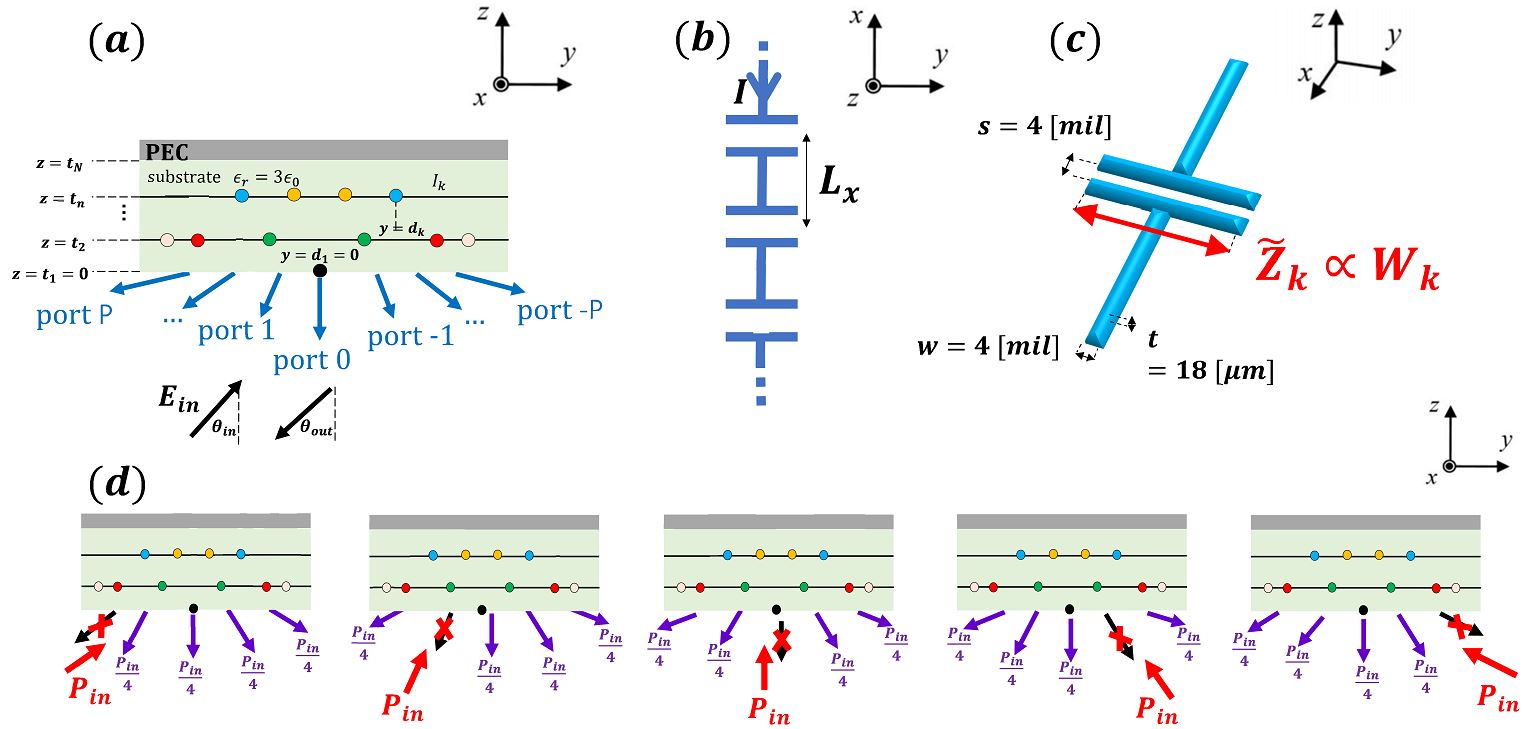}
\caption{(a) Physical configuration of {a} symmetric $M=(2P+1)${-}port PEC-backed PCB MG diffuser. Front view of the MG, {featuring $K$} loaded {copper} strips distributed within the medium {in the planes $z=t_n$,} with horizontal and vertical offsets $d_k$ and $h_k$, respectively{, with respect to the reference meta-atom at $d_1=h_1=0$}. The structure i{s} excited from below by a plane wave{, with an angle of incidence $\theta_\mathrm{in}=\theta_p$ associated with one of the input ports $p=-P,...,P$}. {T}he {denoted} out{put} ports correspond to the {propagating Floquet-Bloch modes scattered from the MG towards the angles {$\theta_q$ (see also Section \ref{subsec:multichannel_MG})}}. (b) Loaded wire meta-atom (top view). The loads repeat periodically with separation distance $L_x=\lambda/10{ \ll \lambda}$ along the $x$ axis, effectively forming a distributed per-unit-length impedance ${\tilde{Z}_k}$. (c) Load detailed geometry. The width $W_k$ of the copper printed capacitor controls the effective distributed impedance ${\tilde{Z}_k}$. (d) {Multi-angle diffusive RCS reduction} functionality illustration for {a $5$-port {multichannel MG [corresponding to panel (a) with $P=2$]}}. For each excitation scenario, the retroreflection vanishes and the {incident} power {is split} equally between the remaining {$4$} ports.}
\label{fig:configuration}
\end{figure*}

To address these issues, we present herein an alternative approach, utilizing the concept of metagrating{s} (MG) {to devise effective diffusive covers with broad acceptance angle}. Similar to MSs, MGs are composed of small polarizable elements, arragned periodically; however, in constrast to MSs, the meta-atoms in MGs are sparsely distributed \cite{Radi2022MetagratingsManipulation}. Due to their sparsity, {{MGs} are not governed by homogenization; instead, to design them, the detailed} interactions between the meta-atoms {are considered based on a suitable analytical model, tying the available geometrical degrees of freedom (DOF) to the scattered fields. Subsequently, by judicious tailoring of the meta-atom distribution and their detailed dimensions, the scattering from individual elements can be engineered to form a} desired interference pattern. This yields {efficient high-fidelity practical designs}, avoiding {the} extensive full-wave optimization associated with MS {synthesis while featuring simpler and easier-to-fabricate layouts} {\cite{Sell2017Large-AngleGeometries,RaDi2017Metagratings:Control, Memarian2017Wide-band/angleResonances, Wong2018PerfectMetasurface}}. 

Indeed, in recent years, we have developed a {semianalycial} scheme to synthesize multilayer printed-circuit-board (PCB) MGs {based on} capacitively-loaded conducting wires {as meta-atoms} [Fig. \ref{fig:configuration}{(a)-(c)}], {demonstrating the ability to manipulate beams at microwave frequencies in versatile manners with high efficiencies \cite{Epstein2017UnveilingAnalysis, Rabinovich2018AnalyticalReflection, Rabinovich2019ExperimentalMetagratings, Rabinovich2020ArbitraryMetagratings, Rabinovich2021NonradiativeMetagratings}. {Similar configurations have been subsequently used in a wide variety of scenarios, allowing diverse static and dynamic beamforming \cite{Popov2018ControllingMetagratings, Popov2020BeamformingDemonstration, Casolaro2020DynamicMetagratings, Popov2021NonlocalManipulations, Xu2021AnalysisApproach}, enhancing waveguide systems \cite{Killamsetty2021MetagratingsExperiment, Biniashvili2022EliminatingMethodology}, and alternative antenna devices \cite{Popov2020ConformalManipulation, Xu2022ExtremeOptimization, Kerzhner2022Metagrating-assistedApplications}}.

To harness this concept to obtain broad-angle reduction of both monostatic and bistatic RCS, we utilize a multichannel scattering perspective {\cite{Asadchy2017FlatReflectors, Yashno2021Multichannel}}. {Since MGs are periodic composites, they must} compl{y} with the Floquet-Bloch (FB) theorem{. In other words,} when illuminated by a plane wave, {real power} {can only be} scattered {into} a finite set of discrete directions {(associated with the various FB modes)}, determined by the angle of incidence and the period length.
Accordingly, we describe the MG as a multiport network, where the in{put} and out{put} channels are defined by the angles of incidence and the corresponding scattering angles, respectively {\cite{Asadchy2017FlatReflectors,Wang2020IndependentDevices}} [Fig. \ref{fig:configuration}{(a)}]. We leverage this method to design a MG, {which} for {\emph{each}} excitation scenario{,} evenly scatters the incident power between the out{put} channels, while specifically suppressing the {coupling to the} retroreflection channel [Fig. \ref{fig:configuration}{(d)}]; thus, {wide-angle} monostatic and bistatic RCS reduction {is achieved} {within} a single {passive} device. 

Specifically, {to design a suitable multielement multilayer MG,} we {harness the model} presented in {\cite{Epstein2017UnveilingAnalysis,Rabinovich2018AnalyticalReflection,Rabinovich2020ArbitraryMetagratings}}, {invoking first the} superposition principle to evaluate separately the {scattering off the given PCB dielectric stack} in the absence of the {MG} grid (external fields), and the {contribution of the} secondary {fields produced by the currents induced on the MG} \cite{S.Tretyakov2003AnalyticalElectromagnetics, RaDi2017Metagratings:Control}. Next, we utilize Ohm’s law to relate these currents to the electric fields {acting on the loaded wires,} {forming a} set of coupled equations {associating the MG properties (wire positions and capacitive loads) with the induced currents and overall FB coupling coefficients}. {Th{is} analytical formulation, in turn, can be used to evaluate} the scattered fields for given MG configuration {and incident beam angle} (the forward problem), {serving} as the basis for the inverse problem {solution:} retrieving the meta-atom constellation and dimensions that would generate the desired interference patterns when illuminated from {{\emph{each} of the \emph{multiple} considered}} excitation angles (input channels){\footnote{{As will be laid out in detail in Section \ref{sec:theory}, this design methodology differs from the one employed in \cite{Epstein2017UnveilingAnalysis,Rabinovich2018AnalyticalReflection,Rabinovich2020ArbitraryMetagratings}, where effectively only the wire coordinates were used directly as DOFs for the synthesis process, and the load impedances were set automatically once these were determined as to satisfy the linear set of constraints on the coupling coefficients (\textit{cf.} Eqs. (17)-(20) of \cite{Rabinovich2020ArbitraryMetagratings}). Herein, we use both the load impedances and the wire coordinates as independent DOFs and rely on inverse solution of the overall (nonlinear) set of constraints (taking into account also multiple excitations), which is essential for achieving the multichannel functionality with a minimal number of meta-atoms per period.}}}. 

Importantly, by {formulating} the desired scattering {requirements for the possible angles of incidence} as a set of constrain{t}s imposed on the structure{'s} scattering matrix ({S}-matrix), it is possible to {evaluate and possibly reduce} the number of DOF{s} {needed to realize the multifunctional MG}. {In general, since we strive to devise a passive and lossless design, reciprocity and power conservation are enforced, identifying the minimal set of independent scattering coefficients.} {However, herein we impose an additional requirement, demanding that the MG configuration would be symmetric. Aligned with the} {broad-angle} RCS reduction scenario{, we show that such a symmetry requirement also reduces the number of overall independent S-parameters, translated {in}to a reduced number of constraints}. In contrast to {other} recently reported multifunctional MG designs{, e.g.} \cite{Radi2018ReconfigurableMetagratings, Yin2019TerahertzCells, Xu2021AnalysisApproach}, {this symmetry-oriented approach we follow} allows dramatic minimization of the required elements per period, leading to a compact and sparse {formations}. This methodology, verified via full-wave simulations and demonstrated experimentally, yields a general and reliable approach {for} designing multichannel MGs for versatile and broad angle beam-manipulation applications{. In particular, it paves the path to enhanced low-profile RCS reduction covers, which are simple to fabricate (sparse) and design (require no full-wave optimization), successfully suppressing both monostatic and bistatic RCS for multiple angles of incidence.}

\section{Theory} \label{sec:theory}
\subsection{{Scattering off a PCB MG (analysis)}} \label{subsec:Formulation}
We consider a $2D$ $(\partial/\partial x = 0)$  $\Lambda${-}periodic structure, composed of $N$ metallization layers embedded in a dielectric substrate of permittivity ${\varepsilon_2} = {\varepsilon_{\mathrm{sub}}}$ bounded within the region $z\in [0,t_N]${, backed by a PEC at $z=t_N$ and} surrounded by a medium {(air, by default)} with {permittivity $\varepsilon_1$} occupying the half-space $z<0$ [Fig. \ref{fig:configuration}(a)].
The wavenumbers and wave impedances for each medium are given, respectively, by $k_i =\omega \sqrt{{\mu_i} \epsilon_i}$, $\eta_i ={\sqrt{\mu_i/\epsilon_i}}$, where $\mu_i$ is the permeability of the $i$th medium; the subscript $i$ refers to the medium where the fields are evaluated $(i=1,2)$. Within a period, $K$ wires with impedances per-unit-length $\tilde{Z}_k$ are distributed among the layers, forming the MG. To maximize the available DOF{s}, the wires can be vertically and horizontally offset with respect to one another, and we denote the $k$th wire position as $(y,z)=(d_k,h_k)$, where $d_k \in (-\Lambda/2, \Lambda/2)$, $h_k \in {\{}t_1, t_2,\ldots , t_{N-1} {\}}$. The bottom metal layer is situated at the substrate-air interface, defined as the $z=t_1=0$ plane.

The structure is excited from below by a transverse electric (TE) {polarized} $(E_z=E_y=H_x=0)$ plane wave, $E_x^{\mathrm{inc}} (y,z) = E_{\mathrm{in}} e^{-jk{y}\sin{\theta_{\mathrm{in}}}-jk{z}\cos{\theta_{\mathrm{in}}}}$, with amplitude $E_{\mathrm{in}}$ and angle of incidence $\theta_{\mathrm{in}}$. According to the FB theorem, the scattered fields {may couple only to a discrete} set of FB modes{, with a finite number of them being propagating (and the rest evanescent)}. {Specifically,} the transverse and longitudinal wavenumbers of the $m$th {FB} mode in the $i$th medium are determined by the period length and the incidence angle via \cite{S.Tretyakov2003AnalyticalElectromagnetics}
\begin{eqnarray} \label{eq:wavenumbers}
\begin{array}{l l}
k_{t_m} = k_1 \sin {\theta_{\mathrm{in}}} + \frac{2\pi m}{\Lambda}, & \beta_{m,i} =\sqrt {k_i^2 - k_{t_m}^2 } 
\end{array}
\end{eqnarray}
with the square{-}root {branch chosen such that $\Im{\{}\beta_{m,i}{\}}\leq 0$ }{to satisfy the radiation condition}.

To find the{se} scattered fields, {w}e follow the formulation presented {in} \cite{Rabinovich2020ArbitraryMetagratings}, {applying it to} the specific case {considered herein,} of a metal-backed
multielement MG with a single dielectric substrate {material}. {Correspondingly, for a given MG configuration, this is achieved by ({i}) evaluating the scattered fields in the absence of the MG grid (external fields); ({ii}) evaluating the secondary fields produced by the currents induced on the MG wires; {and} ({iii}) assessing the actual amplitudes of these induced currents and summing over all field contributions.} 

\subsubsection{{External field contribution}}
\label{subsubsec:external_fields}
As in \cite{Rabinovich2020ArbitraryMetagratings}, due to the {field} discontinuity caused by the presence of the wires{,} we divide the problem {domain} into $N-1$ regions, where the $n$th region is confined {between} the $z=t_{n-1}$ and $z=t_n$ metalization layers. {However, since} the PCB MG herein features a single type of dielectric, for {calculating} the external field it is sufficient to treat separately only two spatial sections; specifically, we distinguish between the fields in the observation region $(n=1)$ and within the dielectric {substrate} $(n>1)$. 

{Consequently,} the external field {(i.e., in the absence of the MG wires)} in the $n$th region can be written as a sum of forward and backward propagating plane waves {adhering Snell's law} 
\begin{equation}  \label{eq:ExtFields}
E^{\mathrm{ext}}_{n} (y,z)= A^{\mathrm{ext}}_{0,n}e^{-jk_{t_0}y - j\beta_{0,i}z} + B^{\mathrm{ext}}_{0,n}e^{-jk_{t_0}y + j\beta_{0,i}z} 
\end{equation} 
where the subscript $i$ refers to
the medium {of the $n$th layer} ($i=1$ for $n=1$ and $i=2$ for $n>1$)  [\textit{cf.} \eqref{eq:wavenumbers}].  The amplitudes $A^{\mathrm{ext}}_{0,n}$ and $B^{\mathrm{ext}}_{0,n}$ are calculated by imposing the relevant boundary conditions{, namely,} continuity of the tangential fields at the air–dielectric interface $(z=0)$ and vanishing of the tangential electric field at the PEC interface $(z=t_N)$, leading to  
\begin{equation}  \label{eq:AB_Ext}
\!\!\!\!\!\!
\begin{array}{l l}
A^{\mathrm{ext}}_{0,1}\!=\!E_{\mathrm{in}} &  \!\! B^{\mathrm{ext}}_{0,1}\!=\!E_{\mathrm{in}}\frac{\Gamma_{0}+ e^{-2j\beta_{0,2}t_N}}{1+\Gamma_{0}e^{-2j\beta_{0,2}t_N}} \\ 
 A^{\mathrm{ext}}_{0,n>1}\!=\! E_{\mathrm{in}} \frac{1+ \Gamma_{0}}{1-\Gamma_{0}e^{-2j\beta_{0,2}t_N}}     &\!\! B^{\mathrm{ext}}_{0,n>1}\!=\! A^{\mathrm{ext}}_{0,n>1} e^{-2j\beta_{0,2}t_N}  
\end{array}
\end{equation}
where the local reflection coefficient {is} defined as {$\Gamma_{m}=\frac{Z_{m,2}-Z_{m,1}}{Z_{m,2}+Z_{m,1}}$}, with the TE wave impedance of the $m$th mode in the $i${th} medium {being} $Z_{m,i}=\frac{k_i \eta_i}{\beta_{m,i}}$. 

\subsubsection{{Grid{-}induced field contribution}}
\label{subsubsec:grid_fields}
Once excited by the incident field, currents would be induced on the MG wires, giving rise to secondary fields. To evaluate these grid-originated fields, we superimpose the contributions of the various wires in the MG, acting as {a} $\Lambda$-periodic array of electric line sources. Specifically, the electric field in the $n$th layer {due to the} (yet to be evaluated) current{s} $I_k$ {developing on the} $k$th wire {in each period} can be written as an infinite series of FB modes
\begin{equation}  \label{eq:Ek_n}
E^{(k)}_n(y,z)= \sum_{m=-\infty}^{\infty} E^{(k)}_{m,n}(y,z)
\end{equation} 
where the field corresponding to the $m$th mode can{, once more,} be expressed as a sum of forward and backward propagating plane waves
\begin{equation}  \label{eq:Ek_mn}
E^{(k)}_{m,n}(y,z)= A^{(k)}_{m,n}e^{-jk_{t_m}y - j\beta_{m,i}z} + 
B^{(k)}_{m,n}e^{-jk_{t_m}y + j\beta_{m,i}z}.
\end{equation} 
To find the amplitudes $A^{(k)}_{m,n}$ and $B^{(k)}_{m,n}$, we use the recursive formalism scheme described in{ \cite{Epstein2013RigorousDiodes, Rabinovich2020ArbitraryMetagratings}}, applying the source conditions (tangential field discontinuity) due to the current-carrying wires in the configuration while considering multiple reflections {within} the grounded dielectric substrate; for brevity, we provide {here only the final results.} 

{Referring to the specific configuration considered herein [Fig. \ref{fig:configuration}(a)],} we {once again} distinguish between the wires positioned at the air-dielectric interface (external wires), and wires located within the substrate (internal wires). For the external wires, we evaluate separately the coefficients in {two} regions: above the wire $(1<n<N)$, and below {it} ($n=1$) 
{{\begin{equation}  \label{eq:AB_extWires}
\!\!\! \begin{array}{l l}
A^{(k)}_{m,1}= 0 & \!\!\!\!\!\!\!\!\!\!\!\!\!\!\!\! B^{(k)}_{m,1}=-\frac{Z_{m,1}I_k}{2 \Lambda}  \frac{\Gamma_{m}+ e^{-2j\beta_{m,2}t_N}}{1+\Gamma_{m}e^{-2j\beta_{m,2}t_N}} \\
A^{(k)}_{m,n\!>\!1}\!\!=\!\!\frac{Z_{m,1}I_k}{2 \Lambda}  \frac{-(1+ \Gamma_{m})}{1- \Gamma_{m}e^{-2j\beta_{m,2}t_N}} &
\!\!B^{(k)}_{m,n\!>\!1}\!\!=\!\!A^{(k)}_{m,n>1} e^{-2j\beta_{m,2}t_N}.
\end{array}
\end{equation}} }
\\
For the internal wires located at some $z=t_{n_k}$, {three} regions are taken into account {separately}: above the wire $(n_k<n<N)$, below the wire within the substrate $(1<n\leq n_k)$, and below the MG $(n=1)$
{{\begin{equation}  \label{eq:AB_intWires}
\!\!\!\!\!\!\!\!\!\!\!\!\!\!\!\!\!\!\!\! \begin{array}{l}
A^{(k)}_{m,1}= 0  \\
A^{(k)}_{m,1<n\leq n_k}= -B^{(k)}_{m,1<n\leq n_k} \Gamma_{m}   \\
A^{(k)}_{m,{n>n_k}}=  
\frac{-Z_{m,2}I_k}{2 \Lambda}  \frac{e^{j\beta_{m,2}t_{n_k}} - \Gamma_{m}e^{-j\beta_{m,2}t_{n_k}}}  {1+\Gamma_{m}e^{-2j\beta_{m,2}t_N}}; \\
\\
B^{(k)}_{m,1}= B^{(k)}_{m,1<n\leq n_k} (1-\Gamma_{m}) \\
B^{(k)}_{m,1<n\leq n_k}= \frac{-Z_{m,2}I_k}{2 \Lambda}  \frac{e^{j\beta_{m,2}t_{n_k}} + e^{-2j\beta_{m,2}t_N}}  {1-\Gamma_{m}e^{-2j\beta_{m,2}t_N}} \\
B^{(k)}_{m,{n>n_k}}=A^{(k)}_{m,n>n_k} e^{-2j\beta_{m,2}t_N}.
\end{array}
\end{equation} }}

Subsequently, for a given {MG} configuration and {given induced currents} $I_k$, the total field in the $n$th layer can be deduced by summing the external field {\eqref{eq:ExtFields}} {with the MG contribution {\eqref{eq:Ek_n}}, reading} 
\begin{equation} \label{eq:E_n_tot}
E^{\mathrm{tot}}_n (y,z)= E^{\mathrm{ext}}_n{(y,z)} + \sum_{k=1}^K \sum_{m=-\infty}^{\infty} E^{(k)}_{m,n} (y,z).    
\end{equation}

\subsubsection{{Evaluation of induced currents}}
\label{subsubsec:induced_currents}
To {enable actual evaluation of \eqref{eq:E_n_tot},}
we {need to} asses{s} the currents {$I_k$} induced on the {loaded} wires {in the given configuration when illuminated by $E^{\mathrm{ext}}_n$}. {To this end, we utilize Ohm's law, relating} the total field applied on a wire to the current flowing through it via the distributed load impedance {$\tilde{Z}_k$ [Fig. \ref{fig:configuration}(c)]}. Explicitly, for the $k$th wire, {this yields} \cite{S.Tretyakov2003AnalyticalElectromagnetics}
\begin{align}  \label{eq:OhmLaw}
I_k \tilde{Z}_k &= E^{\mathrm{ext}}_{n_k}(d_k,h_k)+ \sum_{q\neq k} \sum_{m=-\infty}^{\infty} E^{(k)}_{m,n}(y,z) \nonumber \\ 
&+ \sum_{m=-\infty}^{\infty} E^{(k)}_{m,n_k} (y\rightarrow d_k, z\rightarrow h_k), 
\end{align}
where the first term of the RHS represents the excitation field in the absence of the MG, the second term corresponds to the fields produced by all wires other than the wire itself at its position, and the last term corresponds to the fields produced by the wire itself, forming together the total field {acting} on the $k$th wire. 

The first and second terms are calculated directly by substituting the reference wire position {in}to \eqref{eq:ExtFields}-\eqref{eq:AB_intWires}. For the third term, due to the singularity of the Hankel function at the origin{,} it is not possible to use the pre-calculated terms for the grid induced fields \eqref{eq:Ek_n}-\eqref{eq:AB_intWires}. {Instead}, we follow the technique presented in \cite{Rabinovich2018AnalyticalReflection, Rabinovich2020ArbitraryMetagratings}, and write the $k$th wire self{-}induced field as {a} summation of the field generated by the reference wire on its shell (using the flat wire approximation \cite{S.Tretyakov2003AnalyticalElectromagnetics}), and the field produced by the other strips at the position of the reference wire (interpreted as a series of image sources){, yielding}  
\begin{align}  \label{eq:Eself}
&E^{(k)}_{\mathrm{self}} = \nonumber \\
&\,-\!\frac{I_k}{2}\!\left\{\textstyle\! \frac{\eta_{i}}{\Lambda \cos{\theta_{\mathrm{in}}}}\! +\! \frac{k_{i} \eta_{i} j}{\pi} \!\!\left[ \log{\frac{2\Lambda}{\pi w}} \!+\! \frac{1}{2}\!\!\!\!\!\!  
\sum\limits_{\substack{\footnotesize m=-\infty \\ m\neq 0 }}^{\infty} \!\!\left(\frac{2\pi}{j \Lambda\beta_{m,i}}\!-\!\frac{1}{|m|}\right) \!\!\right]  \!\!\right\} \nonumber \\ 
&\,+\!\!\!\!\!\sum\limits_{m=-\infty}^{\infty}\!\!\!\!\! \left\lbrace\textstyle \frac{k_{i}\eta_{i}}{2\Lambda\beta_{m,i}} I_k + A^{(k)}_{m,n_k}e^{-j\beta_{m,i}h_k} + B^{(k)}_{m,i}e^{j\beta_{m,i}h_k} \right\rbrace,
\end{align}
$w$ being the {copper trace} width [Fig. \ref{fig:configuration}(c)]. 

Since the forward and backward grid-induced field amplitudes {of} \eqref{eq:Ek_mn} are linearly proportional to the currents, \eqref{eq:OhmLaw} can be rewritten as
\begin{eqnarray}  \label{eq:ohmLaw_MatrixForm}
I_k \tilde{Z}_k= E^{\mathrm{ext}}_{n_k}(d_k,h_k) + \sum_{q\neq k} \zeta^{(k)}_{q} {I_q} + \zeta^{(k)}_{\mathrm{self}} I_k
\end{eqnarray} 
where the coefficients $\zeta^{(k)}_{q}$ and $\zeta^{(k)}_{{\mathrm{self}}}$ can be directly extract{ed} from \eqref{eq:Ek_mn}-\eqref{eq:AB_intWires} and \eqref{eq:Eself}, respectively. Thus, the currents can be calculated by a simple matrix inversion ${\mathbf{I_{K\times 1}}= \mathbf{\Psi_{K\times K}}^{-1} \mathbf{E_{K\times 1}^{\mathrm{ext}}}}$, where {$\mathbf{I_{K\times 1}}$} is a vector {composed} of the induced current{s} $I_k$; {$\mathbf{E_{K\times 1}^{\mathrm{ext}}}$}
is {the excitation} vector{, containing} the external field {values} in the {wire} position{s $E^{\mathrm{ext}}_{n_k}(d_k,h_k)$}; and 
{the matrix} {$\mathbf{\Psi_{K\times K}}$} is defined by
\begin{equation}  \label{eq:PsiMatrix}
\mathbf{\Psi}= \begin{pmatrix}
\tilde{Z}_1-\zeta^{(1)}_{\mathrm{self}} & -\zeta^{(1)}_{2}  & \ldots & -\zeta^{(1)}_{K}\\
 -\zeta^{(2)}_{1} & \tilde{Z}_2-\zeta^{(2)}_{\mathrm{self}} &\ldots & -\zeta^{(2)}_{K}\\
 \vdots & \ldots &  \ddots &  \vdots\\
 -\zeta^{(K)}_{1} & \ldots &  -\zeta^{(K)}_{K-1} & \tilde{Z}_K-\zeta^{(K)}_{\mathrm{self}} &\\
\end{pmatrix}
\end{equation} 

\subsubsection{{Evaluation of load impedances}}
\label{subsubsec:load_impedances}
We should recall at this sta{g}e that the MG loaded-wire meta-atoms are physically realized using copper traces in a conventional PCB configuration, featuring printed capacitors as loads {[}Fig. \ref{fig:configuration}{(a)-(c)]}. Thus, to enable assessment of $\mathbf{\Psi}$ of \eqref{eq:PsiMatrix}, evaluate the various currents $I_k$, and subsequently solve the MG scattering problem, it is required to relate the effective distributed impedance $\tilde{Z}_k$ of the various loads to the respective printed capacitor widths $W_k$ used in practice. Following semianalytical formulas developed in previous work, we approximate the reactive part of $\tilde{Z}_k$ {as} $\Im \{ \tilde{Z}_k\}= -2.85K_{\mathrm{corr}}/(\omega L_x W_k {\varepsilon_{\mathrm{eff}{,r}}})$ \cite{K.C.Gupta1996MicrostripSlotlines}, where $W_k$ is the capacitor width [Fig. 1(c)], $L_x$ is the periodicity along the $x$-axis [Fig. 1(b)], and ${K}_{\mathrm{corr}}$ is a frequency-dependent correction factor (estimated as ${K}_{\mathrm{corr}}=0.947$[mil/fF] for $w=s=4$ mil at the working frequency of $f= 20$ GHz {used herein} \cite{Rabinovich2020ArbitraryMetagratings}{)}. The effective {relative} permittivity for the external wires is given by ${\varepsilon_{\mathrm{eff}{,r}}= \frac{\varepsilon_1+\varepsilon_2}{2{\varepsilon_0}}}$ {for meta-atoms at the dielectric-air interface ($n=1$)}, and ${\varepsilon_{\mathrm{eff}{,r}}=\varepsilon_2{/\varepsilon_0}}$ for the internal wires {($n>1$)}{, where $\varepsilon_0$ is the vacuum permittivity}. The resistive part of the distributed impedances, related to the conductor (copper) losses, {{was} estimated using skin depth and cross section considerations as} $\Re \{ \tilde{Z}_k\} \approx 14.5\times 10^{-3} [\eta/\lambda]$ { for the same operating conditions} \cite{Epstein2017UnveilingAnalysis}.

This completes the {\emph{forward}} problem analys{i}s{:} for {given} incident plane wave and MG parameters (i.e.{,} the wire positions $(y,z)= (d_k,h_k)$ and the capacitors widths $W_k$), {one may follow the above formalism to} deduce the distributed impedances from the capacitor widths, evaluate the currents from \eqref{eq:PsiMatrix} {with \eqref{eq:ExtFields}, \eqref{eq:AB_Ext}, \eqref{eq:Ek_mn}-\eqref{eq:AB_intWires}, \eqref{eq:Eself}, and \eqref{eq:ohmLaw_MatrixForm},} and calculate the fields {scattered off the multilayer MG towards the observer via \eqref{eq:E_n_tot} with $n=1$.} 

\subsection{{Multichannel diffusive MG (synthesis)}}
\label{subsec:multichannel_MG}
{Once the analytical model relating a given MG configuration to the fields scattered off it when illuminated from a given $\theta_\mathrm{in}$ is established as in Section \ref{subsec:Formulation} above, we may proceed to tackle the synthesis problem at hand, defining suitable constraints on the coupling coefficients to the various reflected FB modes (for multiple excitations simultaneously in our multichannel scenario), as to facilitate the desired functionality.} 

\begin{figure*}[ht]
\centering
\includegraphics[width=\textwidth , keepaspectratio]{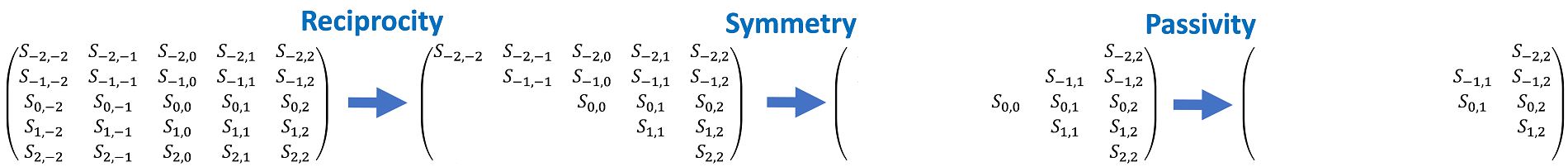}
\caption{{Identification of the} independent {scattering parameters {in}} {a} $5\times 5$ $(P=2,M=5)$ S-matrix, harnessing reciprocity, symmetry{,} and passivity {properties} of the configuration {to yield a compacted set of constraints (see Section \ref{subsec:multichannel_MG})}.}
\label{fig:Smat_constrains}
\end{figure*}

As mentioned in {Section} \ref{sec:introduction}, we wish to exploit symmetry {properties of the problem} to {reduce the overall number of} constrain{t}s {(leading to simpler designs with fewer DOF{s})}. To {obtain} a symmetric {multichannel} network ({about} $\theta= 0^{\circ}$), we {choose} the {$2P+1$ angles of incidence spanning the effective acceptance angle range} to {be} {$\theta_\mathrm{in}\in{\Theta\triangleq}\left\lbrace\arcsin(p\lambda/\Lambda)\,|\,p=-P,...,P\right\rbrace$} {[}Fig. \ref{fig:configuration}{(a),(d)]}. {The advantage of such a choice, beyond manifesting the symmetry of the scattering scenario, is that} for each of th{ese} incidence angles, the propgating scattered FB modes will {be reflected to angles of the same set {$\theta_\mathrm{out}\in{\Theta}$.}} {This would} form a symmetric $M=(2P+1)${-port} network \cite{Asadchy2017FlatReflectors}, where the number of channels (propgating modes) is set by $\Lambda$ via $P=\lfloor  \Lambda/ \lambda \rfloor$ \cite{Rabinovich2020ArbitraryMetagratings}{, promoting a broad-angle response}.

Following this {idea}, and along the lines of \cite{Rabinovich2020ArbitraryMetagratings}, we can {calculate} the {amplitude of the} field {$E_{q,p}$} scattered {towards} the $q${th} port when the MG is excited {from} the $p${th} port {using \eqref{eq:E_n_tot} with $n=1$}{, reading} 
\begin{equation}  \label{eq:Eqp}
E_{q,p}= B^{\mathrm{ext}}_{-p{,}p{;}1} \delta_{q,-p} + \sum_{k=1}^K B^{(k)}_{q{,}p{;}1}.
\end{equation}
where $B^{(k)}_{q{,}p{;}1}$ is the downwards propagating wave amplitude $B^{(k)}_{m,n}$ of \eqref{eq:Ek_mn} in Section \ref{subsec:Formulation}, evaluated in the observation region $n=1$ for a scenario in which $\theta_\mathrm{in}=\theta_p$, and the FB mode order{\footnote{{In contrast to \cite{Rabinovich2020ArbitraryMetagratings}, we use port indices $q$ and $p$ for the scattering field amplitudes, the interpretation of which in terms of incidence/reflection angles is fixed for all excitation scenarios considered herein, rather than using directly the FB mode orders $m$, whose associated scattering angle{s} vary with the angle of incidence. For instance, the specular reflection would always be associated with the fundamental $m=0$ FB mode, while in practice it could refer to different ports (different reflection angles) for different excitations. Naturally, to address systematically the multichannel device presented herein, we found the former notation scheme to be more suitable.}}} $m$ corresponds to the output angle defined by $\theta_q=-\arcsin(k_{t,m}/k_1)$; $B^{\mathrm{ext}}_{-p{,}p{;}1}$ is the amplitude of the reflected external field {$B^{\mathrm{ext}}_{0,1}$} of \eqref{eq:ExtFields} {for the same excitation scenario $\theta_\mathrm{in}=\theta_p$, which only contributes in case that the $q$th port corresponds to the specular reflection, namely, $q=-p$.} 

{{With} these definitions}, we describe the {multichannel} MG {system} using a $M\times M$ scattering matrix ({S}-matrix), where the {input and output} ports {correspond to the FB modes associated with} the excitation angles $\theta_p{\in\Theta}$ and the scattering angles $\theta_q{\in\Theta}$, respectively. {Specifically, the fraction of power coupled to the $q$th port by the MG when excited from the $p$th port can be evaluated via the corresponding scattering coefficient $S_{q,p}$ using} \eqref{eq:Eqp}{, reading} \cite{Rabinovich2018AnalyticalReflection}
\begin{equation} \label{eq:Smatrix}
    |S_{q,p}|^2=\left| \frac{E_{q,p}}{E_{\mathrm{in}}} \right|^2 \frac{\cos{\theta_{q}}}{\cos{\theta_{p}}}
\end{equation}
{This scattering matrix representation allows us to conveniently stipulate the constraints guaranteeing the desired MG functionality, namely, multi-angle monostatic and bistatic RCS reduction. To achieve these goals simultaneously, we demand that} for each excitation scenario, {coupling to the} retroreflection channel would vanish ($S_{p,p}=0$), and the {incident} power would be uniformly scattered into the other available channels} $( |S_{q,p}|^2= 1/(M-1),\: \forall p\neq q)$. The matrix describing the {\emph{power}} scattered from each port at a given excitation scenario can {thus} be written explicitly as{{\footnote{As known, a 3-port matched network could not be {realized} by a passive and lossless configuration \cite{DavidM.Pozar2011Engineering} {such as the one prescribed in \eqref{eq:Smatrix_explicitly}}; therefore{, we restrict} the number of channels {considered} herein to $M>3$.}}}

\begin{equation} \label{eq:Smatrix_explicitly}
\mathbf{|S|}^2=\begin{pmatrix}
 0 & \frac{1}{M-1} & \ldots & \frac{1}{M-1}\\
 \frac{1}{M-1} & 0 &\ldots & \frac{1}{M-1}\\
 \vdots & \ldots &  \ddots &  \vdots\\
 \frac{1}{M-1} &  \frac{1}{M-1} &\ldots &  0\\
\end{pmatrix}    
\end{equation}

{It is important to} note that {in the synthesis scheme we develop here the constraints are applied \emph{directly} to the nonlinear relations tying the scattering parameters to the MG configuration \eqref{eq:Smatrix}. This is in contrast to the method laid out in \cite{Rabinovich2020ArbitraryMetagratings}, where the desired coupling coefficients {of} the propagating FB modes were first substituted into \eqref{eq:E_n_tot} [or \eqref{eq:Eqp}] to form a linear set of equations, treating the induced currents as the only unknowns. Following this approach {of  \cite{Rabinovich2020ArbitraryMetagratings}}, for given meta-atom distribution, the induced currents (and subsequently the load impedances) are determined in a unique manner through this linear system, an observation that was integrated into the design procedure. {Corre{s}pon{d}ingly, the synthesis method adopted therein} merely required from the MATLAB solver \texttt{lsqnonlin} to find these meta-atom coordinates that would lead to a passive and lossless design under this set of linear equations. However, an implicit requirement for the utilization of such a scheme is that the number of unknowns (induced currents) in the linear system would be equal {to} or larger {than} the number of equations (propagating FB modes), i.e. $K\geq M$, which puts a lower bound on the number of meta-atoms per period {required} for \emph{each} considered functionality (design goal) \cite{Rabinovich2020ArbitraryMetagratings}.} 

{Herein, we relax this bound by allowing more flexibility to the inverse problem solver, {looking} simultaneously for suitable meta-atom locations {\emph{as well as}} suitable passive loads to meet the multiple (multichannel) constraints, without explicitly invoking the linear relations between the currents and the FB modal amplitudes for fixed meta-atom placements. Since in such a general nonlinear solution scheme the meta-atom coordinates serve as legitimate DOF{s} which can be used independently from the induced currents, this results in great benefits in terms of scatterer density (reduced number of meta-atoms per period), especially for multifunctional MGs. An additional benefit of this approach is that it does not require stipulation of the complex FB modal \emph{field} amplitudes; instead, the constraints can be defined in terms of the desired fraction of incident \emph{power} coupled to each mode (channel), which is a far less strict limitation, and may be satisfied with fewer DOF{s} (simpler designs).} 

In fact, we can decrease the number of required {DOFs} even further if we harness the S-matrix properties, which should be inherently satisfied by the MG as we design it, reducing the number of independent constraints in our problem. Specifically, we recall that the MG is composed of {reciprocal,} passive and {(ideally)} lossless elements ({highly-}conducting strips {and low-loss dielectrics}){, and that the ultimate configuration should be symmetric about the $\widehat{xz}$ plane}. {Reciprocity implies that} imposing constrain{t}s over $S_{p,q}$ {immediately determines} $S_{q,p}${, removing the need to explicitly consider the latter (Fig. \ref{fig:Smat_constrains}). Furthermore, the symmetric nature of the MG requires that} $S_{p,q}=S_{-p,-q}$ {[Fig. \ref{fig:configuration}(a){,(d)}],  making it sufficient to apply constraints only on one of these S-parameters. Lastly, power conservation implies that the S-matrix should be unitary, and, in particular, the absolute square of the S-parameters in each row or column should sum up to unity} \cite{DavidM.Pozar2011Engineering}{. In other words, once all the S-parameters but one in a row $q'$ are determined, the power coupled to this last element $p'$ is bound by $1-\sum_{p\neq p'} |S_{q',p}|^2$, and thus need not be constrained by our synthesis procedure\footnote{Since our aim is to reduce the RCS in each observation angle, enforcing an upper bound on the scattering is sufficient.} (Fig. \ref{fig:Smat_constrains}).} {In summary, as illustrated in Fig. \ref{fig:Smat_constrains}}, for {a} $M=2P+1$ port network, the number of independent elements {is reduced by reciprocity, symmetry, and passivity from} the initial $M\times M= (2P+1)^2$ {to $(P+1)^2-(P+1)=P(P+1)$, corresponding to a dramatic reduction in the required constraints.}  

{With the number of independent constraints established, we may proceed to executing the synthesis procedure based on a suitable number of DOF{s} (elements per period and their coordinates). To this end, we use the forward problem analytical formulation of Section \ref{subsec:Formulation} and solve the inverse problem defined by the constraints \eqref{eq:Smatrix_explicitly} using} the MATLAB library function \texttt{lsqnonlin}. Since the problem is nonlinear, different initial values of $(d_k,h_k,W_k)$ provided to the function result in convergence to different solutions. To search for an optimal configuration, we run the function with different random initial values 50 times (overall runtime {$<1$} min {on a standard desktop computer}), and choose out of the resulting 50 sets of MG designs the one with the smallest residual (the square norm of the deviation from the desired S-matrix, as defined in {\eqref{eq:Smatrix_explicitly}}). Since this process provides numerous {options} for {detailed} fabrication-ready {PCB layouts}, we may choose a configuration which will best match practical fabrication constrain{t}s (low profile, layer thicknesses that match commercially available dielectric laminates, etc.) {and proceed towards implementation}.

\section{Results and discussion}
\label{sec:results}

{\subsection{Prototype}}
\label{subsec:prototype_design}

To verify {the developed synthesis} method, we follow the scheme described {in Section \ref{sec:theory}} to design a {multichannel} MG diffuser for {multi-angle monostatic and bistatic} RCS reduction. In particular, we {aim at realizing} a 5-channel MG $(M = 5)$ {at $f = 20$ GHz $(\lambda\approx 15 \mathrm{mm})$, corresponding to an S-matrix \eqref{eq:Smatrix_explicitly} which manifests zero retroreflection, and} equally {divides the incident power among} the {four} remaining channels $(|S_{q,p}|^2= 0.25,\: \forall p\neq q)$. {Since we wish the chosen scattered wave (FB mode) trajectories to span well the reflection angular range $\theta\in(-90^\circ, 90^\circ)$, we set the periodicity to} $\Lambda{=\lambda/\sin(29^\circ)}\approx 31 {\mathrm{mm}}${,} {leading to a set of five propagating FB modes, defining the input and output channel propagation angles as $\theta_0=0^{\circ}$, $\theta_{\pm1}=\pm\arcsin(\lambda/\Lambda)=\pm29^\circ$, and $\theta_{\pm2}=\pm\arcsin(2\lambda/\Lambda)=\pm75.84^\circ$ (Section \ref{subsec:multichannel_MG})}. 

\begin{figure}[t]
\centering
\includegraphics[width=\columnwidth , keepaspectratio]{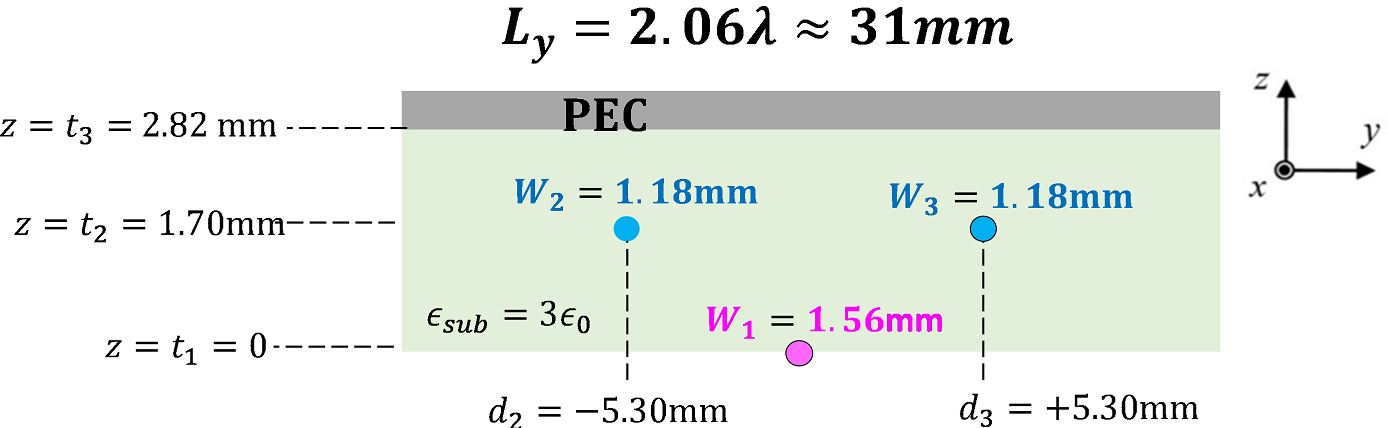}
\caption{{Physical configuration of} the symmetric 5-channel MG prototype {synthesized in Section \ref{sec:results}}{, designed to provide multi-angle monostatic and bistatic} RCS reduction {at $f=20$ GHz} {(}side view, {only} a single period {is shown}).}
\label{fig:protype_configuration}
\end{figure}

We {attempt to design the desired MG prototype based on} a configuration composed of {two metallization} layers $(N=3)$  {of} $0.5$oz {copper traces} ({copper thickness} $t=18\mu$m){,} {embedded within a} Rogers RO3003 $(\varepsilon_{\mathrm{sub}}= 3\varepsilon_0, \tan{\delta}= 0.001)$ {substrate}, {comprising three} meta-atoms per period $(K=3)$. The first {meta-atom (loaded wire)} is {positioned in the origin} at the air-dielectric interface {$(d_1,h_1)=(0,0)$}, {while} the second and third {meta-atoms} are located {at $(d_2,h_2)$ and $(d_3,h_3)$} within the substrate. As {discussed in Section \ref{subsec:multichannel_MG}} above, to reduce the {number of independent constraints and subsequently the} required {number} of DOF{s}, we impose a symmetric configuration, namely, the internal wires are designed with {identical} capacitor widths {$W_2=W_3$} {(identical loads)}, and are positioned {symmetrically about} the origin {($d_2=-d_3$, $h_2=h_3$), as depicted in} Fig. \ref{fig:protype_configuration}. 

{Applying the constraints \eqref{eq:Smatrix_explicitly} via the methodology described in Section \ref{sec:theory} yields a suitable MG design, featuring meta-atoms at $(d_1,h_1)=(0,0)$ and $(d_3,h_3)=(-d_2,h_2)=(0.3507\lambda, 0.1187\lambda)=(5.3\mathrm{mm},1.8\mathrm{mm})$ below the metallic mirror at $z{=t_3}=0.1935\lambda=2.9\,\mathrm{mm}$, with printed capacitors {of widths} $W_1=1.56\,\mathrm{mm}$ and $W_2=W_3=1.18\,\mathrm{mm}$.} Once the geometric parameters are {thus} set, we {may readily evaluate} the S-matrix of the MG by solving the forward problem \eqref{eq:E_n_tot}{-}\eqref{eq:Smatrix} with the finalized $(d_k, h_k, W_k)$, yielding 
\begin{equation} \label{MATLAB_S_mat}
\mathbf{|S|}^2 = \begin{pmatrix}
0.02 & 0.24 & 0.23 & 0.25 & 0.23 \\
0.24 & 0.01 & 0.24 & 0.25 & 0.25 \\
0.23 & 0.24 & 0.04 & 0.24 & 0.23 \\
0.25 & 0.25 & 0.24 & 0.01 & 0.24 \\
0.23 & 0.25 & 0.23 & 0.24 & 0.02 \\
\end{pmatrix}    
\end{equation}
We note that the matrix {agrees} well with the desired scattering goal \eqref{eq:Smatrix_explicitly} {for our case (}$M=5${)}. As expected, the matrix {corresponds to a} reciprocal, symmetric{,} and {passive system}.

To verify the theoretical predictions, a single period of the {prototype} MG was modeled {in} CST Microwave Studio and simulated under periodic boundary conditions. {Since} in practice, the desired layer {configuration} in the fabricated device {was} realized by cascading laminates of standard thicknesses bonded using 2 mil-thick Rogers 2929 bondply (${\varepsilon_\mathrm{sub} = 2.94\varepsilon_0}$ and ${\tan{\delta}} = 0.003$){, we defined this actual structure in the full-wave solver for the final verification purposes}. {Ultimately, also} due to limited availability in real time, the manufactured MG {featured total} thickness of ${t_3}=111$mil $=2.82$ mm, with the meta-atoms positioned {at} $h_1=0$, $h_2=h_3=67$mil $=1.70$ mm{, slightly away from the designated planes. The simulated S-matrix of the corresponding (actual) MG was found to be}
\begin{equation} \label{CST_S_mat}
\mathbf{|S|}^2 = \begin{pmatrix}
0.02 & 0.26 & 0.22 & 0.21 & 0.27 \\
0.26 & 0.00 & 0.21 & 0.28 & 0.24 \\
0.22 & 0.24 & 0.06 & 0.24 & 0.22 \\
0.24 & 0.28 & 0.21 & 0.00 & 0.26 \\
0.27 & 0.21 & 0.22 & 0.26 & 0.02 \\
\end{pmatrix}   
\end{equation}

\begin{figure}[t]
\centering
\includegraphics[width=\columnwidth , keepaspectratio]{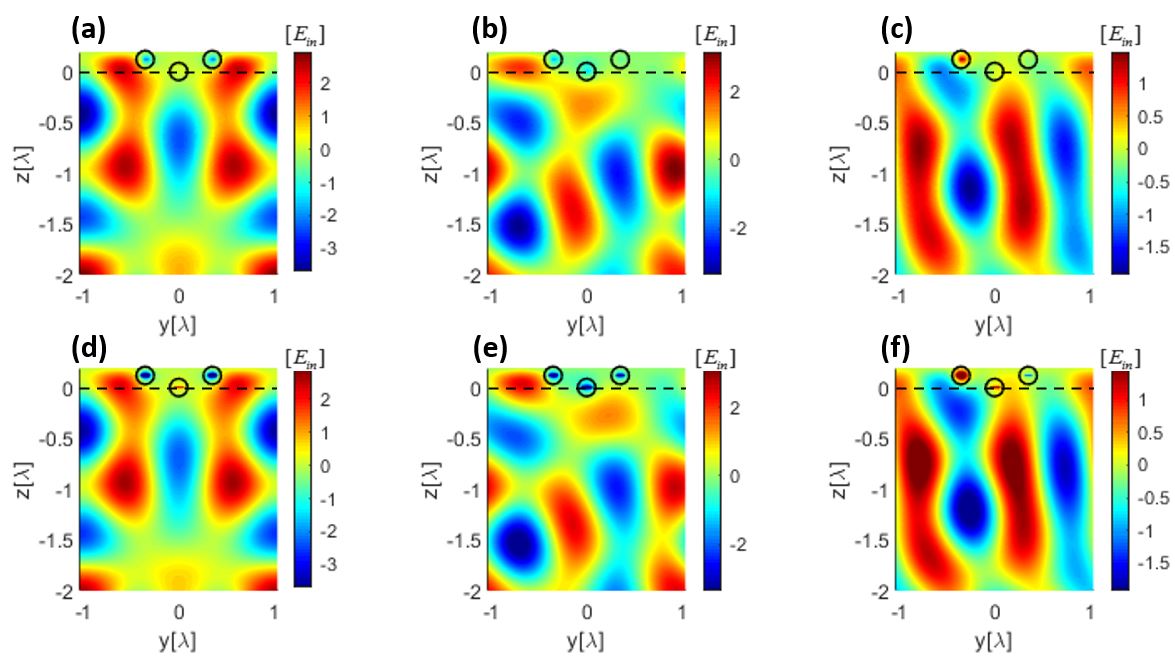}
\caption{Total electric field distributions $\Re\{E_x(y,z)\}$ {for the prototype MG of Fig. \ref{fig:protype_configuration} when illuminated by a plane-wave from} (a){,(d)} $\theta_\mathrm{in}=0^\circ$, (b){,(e)} $\theta_\mathrm{in}=29^\circ$, {and} (c){,(f)} $\theta_\mathrm{in}=75.84^\circ${, as predicted by the analytical model} [(a)-(c)] {and as} recorded in full-wave simulation [(d)-(f)]. Meta-atoms are denoted by black circles; substrate/air interface is marked by a dashed line.}.
\label{fig:Efields}
\end{figure}

{The simulated S-mat shows {overall} good agreement with the analytical{ly predicted one} \eqref{MATLAB_S_mat}{;} the {observed minor} deviation{s} can be {attributed to} the {slight} discrepancies {between} the actual fabricated model {and} the theoretical design {discussed in the previous paragraph}. Furthermore,} Fig{.} \ref{fig:Efields} compares the analytically predicted scattered fields to the full-wave simulated ones for {the} three {designated} angles of incidence, namely, $\theta_\mathrm{in}=0^\circ$ {[}(a),(d){]}, $\theta_\mathrm{in}=29^\circ$ {[}(b), (e){]}, and $\theta_\mathrm{in}=75.84^\circ$ {[}(c),(f){]}, revealing excellent agreement. These results further establish the reliability of the analytical model, demonstrating its usefulness for the design of realistic multichannel PCB MGs.

{\subsection{Experiment}}
\label{subsec:experimental_setup}

\begin{figure}[t]
\centering
\includegraphics[width=0.8\columnwidth , keepaspectratio]{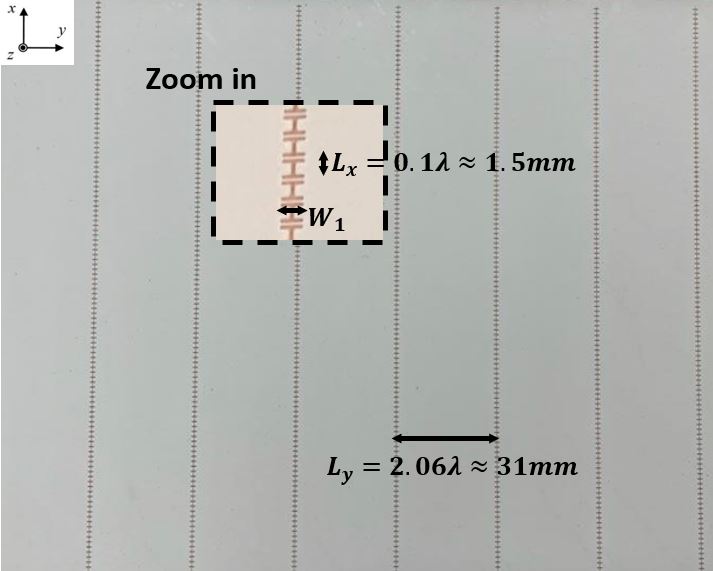}
\caption{The fabricated MG prototype{{, corresponding to} Fig. \ref{fig:protype_configuration}} {(}top view{).} Only the {z=0} layer ({containing} the $k=1$ wire) is visible. {Inset: a close-up view of six capacitive loads on one of the meta-atoms.} }
\label{fig:MG}
\end{figure}

\begin{figure}[t]
\centering
\includegraphics[width=\columnwidth , keepaspectratio]{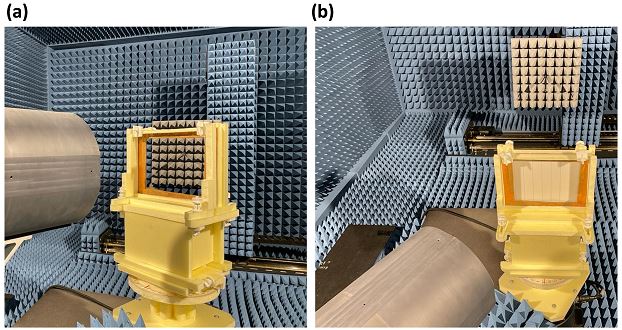}
\caption{{E}xperimental setup {used for characterization of the multichannel MG prototype}. (a) {R}eference measurement using a metallic frame{, enabling} {proper evaluat{ion of} the {effective} excitation power interacting with the MG active area} (no DUT). (b) {Characterization of the} MG under test {in the cylindrical near-field measurement system, conducted under} illuminat{ion} by the Gaussian beam antenna{,} tilted to form a {prescribed} angle of incidence {$\theta_\mathrm{in}$}.}
\label{fig:DUT_REF_expSetting}
\end{figure}

After {this validation of} the theoretical model, and without any further optimization, we used the {corresponding PCB layout} (Fig. \ref{fig:protype_configuration}) to fabricate a $9{''}\times 12{''}$ board ({\textit{PCB Technologies Ltd.}, Migdal Ha'Emek, Israel}){{; the manufactured prototype is} presented in} Fig. \ref{fig:MG}. The {MG device under test (DUT)} was {characterized} in an anechoic chamber at the Technion {using a near-field measurement system ({\textit{MVG/Orbit-FR Engineering Ltd.}, Emek Hefer, Israel}). In the chamber, the MG was} positioned on a foam holder in front of of a Gaussian beam antenna (Millitech, Inc., GOA-42-S000094, focal distance of $196$mm $\approx$ $13\lambda$), illuminating the device with a quasi-planar wavefront {(Fig. \ref{fig:DUT_REF_expSetting}) }. Due to blockage effects presented by the Gaussian beam antenna when scanning {at wide angles}, {we have placed} the MG at a {larger} distance from the Gaussian beam antenna{, corresponding to} $430\,\mathrm{mm}\approx 29\lambda$. {Nonetheless, as can be deduced from previous work} \cite{Rabinovich2020Dual-polarizedReflection}{, the focal region is large enough to provide a sufficiently collimated beam even at this distance, enabling proper characterization of the DUT.}

\begin{figure}[t]
\centering
\includegraphics[width=0.8\columnwidth , keepaspectratio]{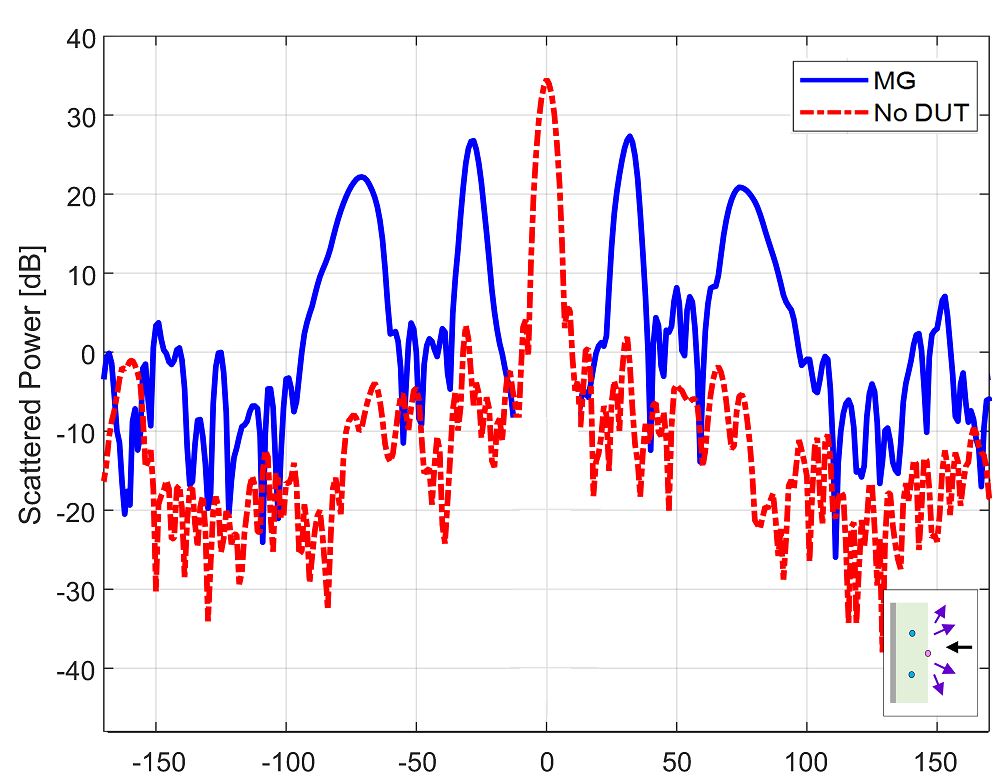}
\caption{Experimentally recorded scattering pattern for normal excitation [in{put} port {$p=0$} in Fig. \ref{fig:configuration}(a)] at the operating frequency of $f = 20$ GHz{. The measurements are conducted once in the absence of the MG} [dash-dotted red{, \textit{cf.} Fig. \ref{fig:DUT_REF_expSetting}(a)]} {and in the presence of the MG} [solid blue{, \textit{cf.} Fig. \ref{fig:DUT_REF_expSetting}(b)], with the former serving as a reference for the input power}.}
\label{fig:Gain_ref_DUT}
\end{figure}

{{To facilitate} the manufacturing process, the MG {circumference was partly sacrificed to accommodate} metallic markings and drilled holes used for accurate alignment during fabrication{, leaving} an active area smaller than the overall $9''\times12''$ board size. Therefore, to properly calibrate our measurements and avoid interaction with these irrelevant scatterers,} we {attached to the foam holder} a metallic frame {covering these markings,} allowing quantification of the {effective} {reference} excitation power {which illuminates the MG area in {reality}} [Fig. \ref{fig:DUT_REF_expSetting}(a)].

{\subsection{Characterization}}
\label{subsec:characterization}
By rotating the foam holder about its axis, the alignment between the Gaussian beam antenna and the MG was tuned such that the DUT scattering properties for different angles of incidence could be probed. In particular, measurements were performed for the excitation angles corresponding to the designated input ports of the multichannel MG [Fig. \ref{fig:DUT_REF_expSetting}(b)]. For each {of these} angle{s} of incidence, a cylindrical near-field measurement {was performed, recording the scattering pattern across} the frequency range {$f\in[18,22]\,\mathrm{GHz}$}. The measurement was conducted by rotating {together} the Gaussian beam {antenna} and the MG (aligned to a certain illumination angle $\theta_{\mathrm{in}}$) around {a near-field} probe {situated at a distance of $850\mathrm{mm}\approx57\lambda$ away from the MG, collecting the scattered fields}. The far field patterns {were} then deduced {by the system's postprocessing software} from the near field measurements using the equivalence principle \cite{C.A.Balanis2012AdvancedElectromagnetics}. 

Figure \ref{fig:Gain_ref_DUT} {shows such representative} scattering pattern{s, recorded} for normal {incidence} {[}{input} port {$p=0$} in Fig. \ref{fig:configuration}(a){]} at the operating frequency of $f = 20$GHz. {As seen, in the absence of the DUT (dashed red curve), a high-gain beam peaking at $\theta=0^\circ$ is recorded in transmission, quantifying the input power effectively interacting with the MG. Once the MG is placed in the holder (solid blue), the expected diffusive scattering is observed, with $M-1=4$ dominant beams. Although these beams peak} around the designated output port angles $\theta_{\pm1}\approx \pm 29^\circ, \theta_{\pm2}\approx \pm 75.84^\circ$, {a closer examination reveals that the exact maxima occur at}  {$\theta_{1}= 31^\circ$, $\theta_{-1}=-28^\circ$, $\theta_{2}= 73^\circ$, {and} $\theta_{-2}= -71^\circ$.} These results lead to two observations. First, the asymmetry in the recorded peak angles implies that the limited (manual) alignment accuracy of the system may introduce angular deviations of up to {around} $1^\circ$ in measurements. Second, for large deflection angles $\theta_{\pm2}$, the main lobe {maxima} shift towards lower angl{ular values, which is attributed to the reduced} effective aperture size {at these near-grazing angles} \cite{Chen2018TheoryRefraction}{.}

{Another aspect that is highlighted by Fig. \ref{fig:Gain_ref_DUT} is the inability of the described (\emph{bistatic})} measurement setup to measure the scattered power for $\theta_{\mathrm{in}}=\theta_{\mathrm{out}}$ {(manifested in the absence of measured data points around $\theta=0$ in the plot when the MG is present). Since in our near-field system the probe and Gaussian beam antenna cannot be aligned to the same angle (otherwise they will physically collide), an alternative method (\emph{monostatic}) should be used to evaluate retroreflection. Specifically, we connect to this end} the {G}aussian beam antenna to a {dedicated} vector network analyzer (4-port Keysight E5080B ENA), functioning as both the transmitter and the receiver {for this measurement. Subsequently, the recorded reflection coefficient} ($S_{11}$) {provides a measure for the amount of retroreflected power. To quantify the fraction of power coupled to this channel, we compare this value with a reference measurement} conducted {with} a planar metal plate {matching} the dimensions of the MG. 

{Finally, for both bistatic and monostatic setups, we} evaluate the power coupled to each {channel by considering the ratio between} the peak measured gain values of the MG scattering patterns $G_{\mathrm{MG}}(\theta_{\mathrm{out}})$ {and} the reference measurement $G_{\mathrm{ref}}(\theta_{\mathrm{in}})$ {quantifying the effective incident power}, taking in account {the differences in the} effective aperture size \cite{Diaz-Rubio2017FromReflectors,Wong2018PerfectMetasurface, Rabinovich2020Dual-polarizedReflection,Rabinovich2020ArbitraryMetagratings}
\begin{equation}
    \eta_{\mathrm{tot}}= \frac{G_{\mathrm{MG}}(\theta_{\mathrm{out}})}{G_{\mathrm{ref}}(\theta_{\mathrm{in}})} \frac{\cos{\theta_{\mathrm{in}}}}{\cos{\theta_{\mathrm{out}}}}{,}
\end{equation} 
{which for the retroreflection case $\theta_\mathrm{in}=\theta_\mathrm{out}$ simply reduces to the ratio of the measured reflection coefficients in the input of the Gaussian beam antenna.}

\begin{figure*}[ht]
\centering
\includegraphics[width=\textwidth , keepaspectratio]{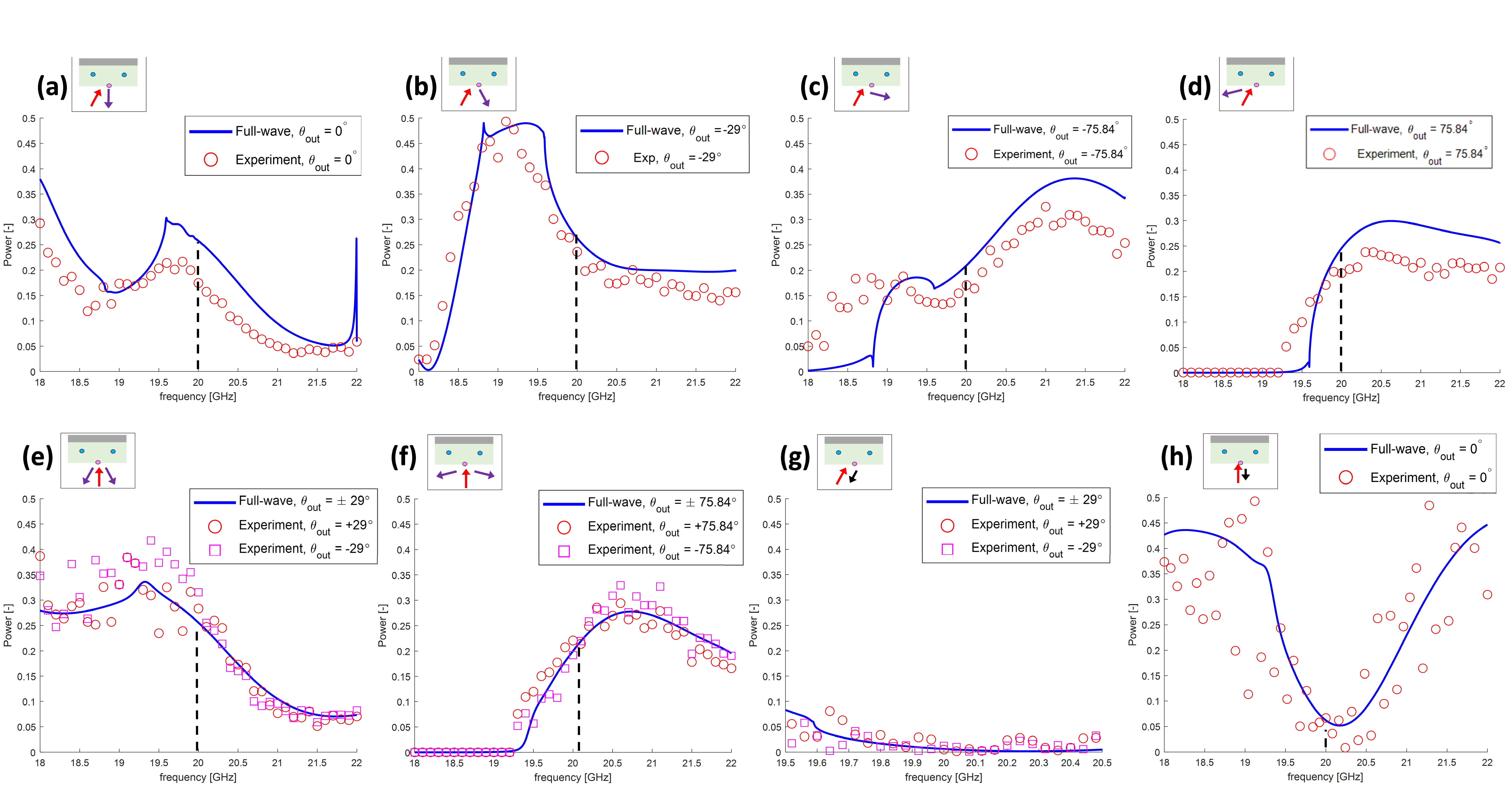}
\caption{{Frequency response of the prototype MG diffuser of Fig. \ref{fig:protype_configuration}.} The fraction of incident power coupled to each {of the output ports} for {two} different {excitation} scenarios {$\theta_{\mathrm{in}}=29^{\circ}$ [(a)-(d),(g)] and $\theta_{\mathrm{in}}=0^{\circ}$ [(e),(f),(h)]} {is plotted} as a function of the frequency. The experimental results ({red} circles and {magenta} squares) are compared with the results obtained via full-wave simulation (solid blue) {for the input and output ports associated with the angles} 
(a) $\theta_{\mathrm{in}}=29^{\circ},  \theta_{\mathrm{out}}= 0^{\circ}${;}
(b) $\theta_{\mathrm{in}}=29^{\circ},  \theta_{\mathrm{out}}= -29^{\circ}$ {;}
(c) $\theta_{\mathrm{in}}=29^{\circ},  \theta_{\mathrm{out}}= -75.84^{\circ}${;}
(d) $\theta_{\mathrm{in}}=29^{\circ},  \theta_{\mathrm{out}}= 75.84^{\circ}${;}
(e) $\theta_{\mathrm{in}}=0^{\circ},    \theta_{\mathrm{out}}= 29^{\circ}$ {(red circles), $\theta_{\mathrm{out}}= -29^{\circ}$ (magenta squares);}
(f) $\theta_{\mathrm{in}}=0^{\circ},    \theta_{\mathrm{out}}= 75.84^{\circ}$ {(red circles), $\theta_{\mathrm{out}}= -75.84^{\circ}$ (magenta squares);}
(g) $\theta_{\mathrm{in}}=29^{\circ},  \theta_{\mathrm{out}}= 29^{\circ}$ {(red circles), and $\theta_{\mathrm{in}}=-29^{\circ},  \theta_{\mathrm{out}}= -29^{\circ}$ (magenta squares)}. 
(h) $\theta_{\mathrm{in}}=0^{\circ},    \theta_{\mathrm{out}}= 0^{\circ}$;
{Insets depict the characterization scenario, while} dashed vertical lines {mark} the operating frequency $f=20$ GHz. }
\label{fig:FreqResonse_fullwaveExp}
\end{figure*}

\begin{figure*}[ht]
\centering
\includegraphics[width=\textwidth , keepaspectratio]{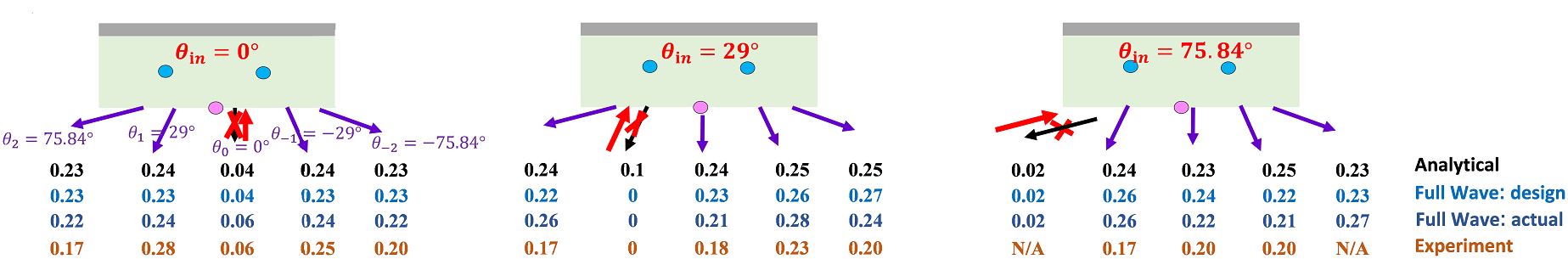}
\caption{{Quantification of the scattering parameters of the prototype MG diffuser of Fig. \ref{fig:protype_configuration}.} The fraction of incident power coupled to each {of the output ports} for all {five considered excitation} scenarios {(input ports) is presented}, {as evaluated at} the operating frequency $f=20$ GHz [due to symmetry, only {three excitation} scenarios {are} presented{, \textit{cf.} \eqref{MATLAB_S_mat}, \eqref{CST_S_mat}}]. The {analytical predictions} (black) are compared with the {scattered power as recorded in full-wave simulations for the optimal design} (light blue) and {the one eventually realized with the available laminate stack} (dark blue), {as well as with the results obtained in} the experiment (orange). Due to the setup limitations, results for the {grazing angle} excitation scenario{s} of ${\theta_\mathrm{in}}= \pm {75.84^\circ}$ were not recorded{; where possible, these were deduced by reciprocity considerations.} }
\label{fig:Smat_TheoryFullwaveExp}
\end{figure*}

Fig{ure} \ref{fig:FreqResonse_fullwaveExp} presents the measured frequency {dependency} of the power coupled to each propagating mode ({output} channels), for illumination from the predefined excitation angles ({input} channels{\footnote{{The measurement setup did not allow proper characterization of the MG when illuminated by large oblique angles of incidence, mainly due to the small effective aperture size for such angles (proportional to $\cos\theta_\mathrm{in}$ \cite{C.A.Balanis2012AdvancedElectromagnetics}{)}. Thus, Fig. \ref{fig:FreqResonse_fullwaveExp} only considers $\theta_\mathrm{in}=0^\circ,\,\pm29
^\circ$, whereas the performance associated with $\theta_\mathrm{in}=\pm75.84^\circ$ is deduced from reciprocity considerations (see discussion {surrounding} Fig. \ref{fig:Smat_TheoryFullwaveExp} later on).}}}), compared {to} the simulation results. The graphs indicate good correspondence between the simulated and experimental{ly} recorded scattering {patterns} in the measured frequency range. {The minor differences observed [mainly in panels (a)-(d)] between theoretical predictions and experimental estimations can be attributed to limited angular alignment accuracy in our setup {(see discussion after Fig. \ref{fig:Gain_ref_DUT})}, which may introduce small errors in the actual angle of incidence with respect to the desired $\theta_\mathrm{in}$; possible fabrication inaccuracies and material parameter tolerances may also contribute to such deviations.} 

The measured frequency response is further affected by the periodic nature of the designed multichannel MG. Indeed, certain effects observed in Fig. \ref{fig:FreqResonse_fullwaveExp} stem from the variation of the particular propagation angle associated with the scattering towards the $p$th port, as dictated by \eqref{eq:wavenumbers}. Specifically, when reducing the operating frequency, higher-order FB modes are driven into the invisible region, eventually becoming evanescent (and thus cannot outcouple real power). This explains the low-frequency behaviour observed in Fig. \ref{fig:FreqResonse_fullwaveExp}(c), (d), and (f), where the scattering towards the presented port abruptly vanishes when a certain "cutoff" frequency is crossed\footnote{In Fig. \ref{fig:FreqResonse_fullwaveExp}(c)-(d), {finite} power is {recorded} at frequencies lower than the theoretical "cutoff" frequency {[}calculated as $f=18.8$ GHz for (c), and $f=19.6$ GHz for (d){]}. This {may be attributed again} to small {alignment errors due to the manual mechanism used in the experiment (see discussion after Fig. \ref{fig:Gain_ref_DUT}), correspondingly causing minor deviations in the} angle of incidence. {Due to the high angular sensitivity for this extreme deflection towards near-grazing angles, even such minor deviations would be sufficient to} shift the "cutoff" frequency below $f=18$GHz {[\textit{cf.}} \eqref{eq:wavenumbers}{]}.}.
Such a crossing {has an impact on} the distribution of power amongst the other FB modes {as well}, as can be seen, for instance, in the substantial growth in the power coupled to the $p=0$ port around $f=19.2$ GHz in Fig. \ref{fig:FreqResonse_fullwaveExp}(h), stemming from the cutoff identified for the same excitation scenario in Fig. \ref{fig:FreqResonse_fullwaveExp}(f). {Similarly abrupt trend changes can be spotted in Fig. \ref{fig:FreqResonse_fullwaveExp}(a) and (b) around the cutoff frequencies $f=18.8$ GHz and $f=19.6$ GHz, associated, respectively, with the FB modes characterized in Fig. \ref{fig:FreqResonse_fullwaveExp}(c) and (d) for the same angle of incidence $\theta_\mathrm{in}=29^\circ$.}

{Another effect of the modal frequency dependency manifested by \eqref{eq:wavenumbers} is related to the retroreflection phenomena. While for $\theta_\mathrm{in}=0$, retroreflection coincides with the specular reflection and thus always directed towards broadside [Fig. \ref{fig:FreqResonse_fullwaveExp}(h)], for other illumination angles (e.g.{,} $\theta_\mathrm{in}=29^\circ$) the multichannel scattering scenario would not include ports with $\theta_\mathrm{out}=\theta_\mathrm{in}$ for measurements outside the designated operating frequency. In other words, significant power could be recorded at $\theta_\mathrm{out}=\theta_\mathrm{in}$ only for a limited range {of frequencies} around the {design working point} $f=20$ GHz (unless $\theta_\mathrm{in}=0$), whereas outside this range coupling towards this direction would not be allowed as per the FB theorem. Following this observation, we present the retroreflection data for $\theta_\mathrm{in}=29^\circ$ only for the range $f\in[19.5, 20.5]$ GHz [Fig. \ref{fig:FreqResonse_fullwaveExp}(e)], in which the deviation of $\theta_\mathrm{out}$ from $\theta_\mathrm{in}$ is still mild such that the retroreflection measurement is still meaningful\footnote{{Recall that for the retroreflection measurement the Gaussian beam is static with respect to the MG, relying on the assumption that the main scattering would be towards $\theta_\mathrm{out}=\theta_\mathrm{in}$ (Section \ref{sec:results}).}}.}

{Overall, it can be seen that the MG performs well around the designated operating frequency, suppressing coupling to the retroreflection mode  while distributing the scattered power amongst the other channels. These properties are further apparent from Fig. \ref{fig:Smat_TheoryFullwaveExp}, quantifying the fraction of power coupled to the various output ports for each of the considered excitations $\theta_\mathrm{in}=0^{\circ}$ (left), $\theta_\mathrm{in}=29^{\circ}$ (center), and $\theta_\mathrm{in}=75.84^{\circ}$ (right) at $f=20$ GHz. For comparison, four sets of data are presented for each scenario. The first row (Analytical) corresponds to the predictions of the analytical model as output by the synthesis procedure {\eqref{eq:E_n_tot}{-}\eqref{eq:Smatrix}}{, in correspondence with \eqref{MATLAB_S_mat}}; the second row (Full-wave: design) presents the scattered power as recorded in full-wave simulations for this chosen design; the third row (Full-wave: actual) indicates the expected effects on the MG performance when the actual laminate stack used for the fabricated prototype is considered in CST{, previously reported in \eqref{CST_S_mat}}; and the fourth row (Experiment) documents the scattered power as measured in the anechoic chamber. As can be clearly seen,} the collected data show {good} agreement between analytically computed, simulated and measured power coupling to each port{. Even the constraints posed by laminate availability did not deteriorate significantly the MG operation, implying a certain robustness against fabrication errors. From a RCS reduction perspective, for all considered angles of incidence in this multichannel scenario,} the measured retroreflection {was found to be} lower than 6\% of the incident power, and the maximal power coupled to a single direction ({output} port) {did not exceed} 28\%. {Hence, the presented results verify the efficacy of the proposed MG in realizing intricate multi-angle scattering management, relying on a PCB fabrication-ready sparse configuration designed using a full-wave-optimization-free methodology.}

{{Before we conclude, we should note that} although the device was designed to equally divide the power between $M-1=4$ ports {only} when excited from one of the $M=5$ designated angles of incidence {$\Theta$}, {it actually reduces the maximal scattering for other excitation angles as well. Specifically,} examining the angular response {by both full-wave simulations and analytically} {over a broader range} reveals that the power coupled to a single direction {(FB mode)} does not exceed 45\% of the incident power, for {all} excitation angles in the range of $|\theta_\mathrm{in}| \leq 80^{\circ}$ excluding a small fragment around $\theta_\mathrm{in}=25^{\circ}$.
Clearly, increasing the periodicity {$\Lambda$} beyond $P\lambda$ for some $P\in\mathbb{N}$ would correspondingly increase the "sampling" resolution of the angular domain with $M=2P+1$ symmetric channels, while at the same time reduce the power fraction scattered to each channel $\propto 1/(M-1)$. Naturally, this would require more DOF{s} (meta-atoms per periods) to meet the increased number of constraints (S parameters); on the other hand, it is expected that if the angular density of the channels is sufficiently high (the angular difference between adjacent angles $\theta_p$ would be sufficiently small){,} the RCS reduction performance would be maintained more evenly across the entire angular range \cite{Wang2020IndependentDevices}.

\section{Conclusion}
\label{sec:conclusion}
In this paper, we introduced a rigorous {semi}analytical method for designing {PCB-compatible} multichannel MG {diffusers, enabling reduction of monostatic and bistatic RCS for multiple angles of incidence simultaneously}. \textcolor{black}{By {adopting} a symmetric and passive MG {construct}, along with a synthesis procedure that clearly identifies the minimal required number of DOF{s} and {effectively exploits them}, a {highly-}sparse configuration is obtained. Specifically, the {demonstrated prototype} achieves multi-angle functionality spreading over a wide angular range with only {three subwavlength loaded wires} within a two{-}wavelength period, compar{ed to the typical number of} $10-20$ {polarizable particles per period that would be used} in a {conventional (single-functionality)} MS {with the same periodicity}. Moreover, the presented model, verified experimentally, enables producing a complete fabrication{-}ready design without relying on time consuming full-wave optimization. Since the analytical scheme is general, {the methodology can be used, in principle, to exercise} control over an arbitrary number of diffraction modes {(by including additional meta-atoms per period)} and {meet the requirements of} versatile {multifunctional} scattering scenarios. In the {context} of the {explored} RCS reduction problem, {this generality can be harnessed to} further enhance the {diffuser} performance {by increasing the} number of channels, {correspondingly improving the angular response as well as the bistatic RCS reduction}. These results {and conceptual observations establish a reliable} foundation {for} the design of sparse multi-angular metagratings {for a variety of electromagnetic applications}.}

\section*{Acknowledgment}
This research was supported by the Israel Science Foundation (Grant No. 1540/18) as well as the PMRI – Peter Munk Research Institute - Technion. The authors also wish to thank Rogers Corporation for providing part of the laminates used in work, and the Keysight team in Israel for providing the 4-port VNA used for the retroreflection measurements. }

\bibliographystyle{IEEEtran}
\bibliography{references}
\end{document}